\documentclass[a4paper,11pt]{article}
\usepackage{jheppub} 
\usepackage{lineno}
\usepackage{txfonts}
\usepackage{physics}
\usepackage{subcaption}
\usepackage{mathrsfs}
\usepackage{amsmath}
\usepackage{tikz}
\usetikzlibrary{arrows,matrix,positioning}
\usepackage{appendix}

\newcommand{\rn}{Reissner-Nordstr\"om}

\title{Decoherence of quantum superpositions by {\rn} black holes}

\author{Ran Li}

\affiliation{Department of Physics, Qufu Normal University, Qufu, Shandong 273165, China}

\emailAdd{liran@qfnu.edu.cn}

\abstract{Recently, it was shown by Danielson-Satishchandran-Wald (DSW) that for the massive or charged body in a quantum spatial separated superposition state, the presence of a black hole can decohere the superposition inevitably towards capturing the radiation of soft photons or gravitons \cite{Danielson:2022tdw,Danielson:2022sga,Danielson:2024yru}. In this work, we study the DSW decoherence effect for the static charged body in the {\rn} black holes. By calculating the decohering rate for this case, it is shown that the superposition is decohered by the low frequency photons that propagate through the black hole horizon. For the extremal {\rn} black hole, the decoherence of quantum superposition is completely suppressed due to the black hole Meissner effect. It is also found that the decoherence effect caused by the {\rn} black hole is equivalent to that of an ordinary matter system with the same size and charge.}

\begin{document}

\maketitle

\section{Introduction}
\label{sec:intro}

Recently, Danielson, Satishchandran, and Wald (DSW) demonstrated that the presence of black hole Killing horizon can inevitably decohere the quantum spatial superposition of a charged massive particle \cite{Danielson:2022tdw,Danielson:2022sga,Danielson:2024yru}. This observation arises from the fact that soft photons or gravitons emitted by a charged massive particle near the black hole can carry ``which-way" information of the quantum superposition.

This effect has attracted significant attention. In \cite{Gralla:2023oya},  a general formula for the precise decoherence rate was derived for the charged massive particle in the rotating Kerr black hole. However, the calculations were limited to the scalar and electromagnetic analogs. To better understand the thermal nature of the DSW decoherence effect, a model of an Unruh-DeWitt particle detector near the Rindler horizon was investigated in \cite{Wilson-Gerow:2024ljx}, which indicates that the thermal environment can induce steady decoherence on the accelerating detector. In \cite{Biggs:2024dgp}, by modeling black hole as a quantum system at finite temperature, it was also shown that the decoherence phenomenon could also be interpreted in this context. Furthermore, it was analyzed that the decoherence effect caused by black hole horizon is comparable to the analogous effect induced by ordinary matter system \cite{Biggs:2024dgp}. 

Based on the aforementioned works, the decoherence effect of a quantum superposition induced by a black hole can essentially be treated as environment-induced decoherence. This kind of effect has been a subject of intensive research \cite{Zurek:2003zz,Schlosshauer:2003zy}. Decoherence, along with the quantum measurement problem, is closely related to the fundamental interpretation of quantum mechanics. Although on the experimental side, it is difficult to generate spatial quantum superposition states for large masses, there are some valuable explorations on this topic in the optomechanical systems \cite{Carlesso:2019cuh} and in Bose-Einstein condensates \cite{Howl:2020isj}. Therefore, the decoherence of quantum superpositions by black hole is a particularly fascinating effect, and understanding the nature of DSW decoherence may provide intriguing insights into both the quantum nature of black holes and the black hole information problem.

In this work, we study the DSW decoherence effect for a static charged body in the {rn} black holes. Our calculation is limited to the case of electromagnetic radiation from the charged particle in a spatial separated quantum superposition state. By calculating the decohering rate for the current case, it is shown that the superposition is decohered by the low frequency photons that propagate through the black hole horizon. In other words, we provide a detailed calculation demonstrating that the quantum superposition is violated due to the entanglement between the black hole and the charged particle, which arises from the black hole absorbing the entangling photon emitted by the charged particle. We also discuss the case of the extremal {\rn} black hole. It is shown that the decoherence of quantum superposition is completely suppressed due to the black hole Meissner effect \cite{Wald:1974np,King:1975tt,Bicak:1980du,Bini:2008iwy}. At last, we compare the decoherence effect due to the {\rn} black hole with that of the ordinary material system and find that the decoherence effect caused by the {\rn} black hole is equivalent to that of an ordinary matter system with the same size and charge.

This paper is arranged as follows. In Sec.\ref{sec:super_experiment}, we briefly review the setup of the DSW gedankenexperiment for decoherence of quantum superposition. In Sec.\ref{sec:decoh_ent_phot}, we discuss how the decoherence rises from the entangling photon radiation. In Sec.\ref{sec:est_num}, we make an estimation of the entangling photon number in the {\rn} black hole spacetime. In Sec.\ref{sec:decoh_flux}, we calculate the decohering flux and give an exact expression for the decoherence rate. In Sec.\ref{Sec_comparing}, we compare the decoherence effect due to the {\rn} black hole with that of the ordinary material system. The conclusion and the discussion are presented in the last section.

\section{Superposition experiment}
\label{sec:super_experiment}

In this section, we briefly review the basic setup of the DSW gedankenexperiment for the quantum superposition state of an charged massive particle in the static black hole background \cite{Danielson:2022tdw,Danielson:2022sga,Danielson:2024yru}. The setup was also been generalized to the stationary Kerr black hole spacetime in \cite{Gralla:2023oya}.

Consider a charged massive particle in a static black hole spacetime.
Suppose that the particle is initially in the positive direction state of $x-$spin. An experimenter sends the particle through a Stern-Gerlach apparatus oriented in the $z-$ direction. After the process, the particle is in a superposition state of the following form
\begin{eqnarray}\label{superposition_state}
    |\Psi\rangle= \frac{1}{\sqrt{2}}\left(|\psi_L,\uparrow\rangle+|\psi_R,\downarrow\rangle\right)\;,
\end{eqnarray}
where $|\psi_L\rangle$ and $|\psi_R\rangle$ are the spatially separated, normalized states of the charged particle after passing through the apparatus and $|\uparrow\rangle$ and $|\downarrow\rangle$ are the spin-up and spin-down states along the $z-$axis direction. The experimenter holds the stationary superposition state for a long period of time, $T$ and then recombines the particle via a reversing Stern-Gerlach apparatus. After recombination, the particle can, in principle, be measured to verify whether the initial state is recovered. If the coherence of the superposition state \eqref{superposition_state} is preserved in the long period of time $\Delta T$, the spin of the particle will always be found in the initial positive $x-$direction.

In the whole process, it is assumed that there are no external influences on the charged massive particle that could induce the potential decoherence of the superposition. Apparently, it may seem that the decoherence of the quantum superposition must be preserved. However, the electromagnetic field or the gravitational field associated with the charged massive particle must be considered as part of the total system. In other words, the electromagnetic radiation or gravitational radiation emitted by the charged massive particle may provide the ``which way" information of the particle. Assuming that the separation and recombination processes occur adiabatically, the emission of entangling radiation to null infinity can be made negligible. Based on this assumption, it has been analyzed that the coherence of the superposition can be maintained in the Minkovski spacetime \cite{Belenchia:2018szb,Danielson:2021egj}. However, an analysis of the gedankenexperiment in the black hole spacetime shows a dramatically different conclusion. It is unavoidable that low-frequency electromagnetic or gravitational radiation can be captured by the black hole horizon. This radiation effectively disappears from the outside spacetime due to the causal structure of the horizon, akin to an observer behind the horizon performing a ``which-way" experiment for the charged massive particle. Therefore, due to the causal structure of the black hole horizon, the coherence of the superposition states of the charged massive particle is violated inevitably \cite{Danielson:2022tdw}.

\section{Decoherence from the entanglement photons}
\label{sec:decoh_ent_phot}

To perform a detailed analysis, it is therefore convenient to consider the time evolution of the state for the total particle-radiation system. We mainly focus on the electromagnetic radiation case in the present work.

Prior to going through the Stern-Gerlach apparatus, the state of the electromagnetic radiation is assumed to be in the vacuum state $|\Psi_0\rangle$ of the radiation field $\boldsymbol{A}_{\mu}^{\textrm{in}}$, which is defined by subtracting the Coulomb field of the charge from the electromagnetic field as \cite{Danielson:2022sga}
\begin{eqnarray}\label{A_in_def}
    \boldsymbol{A}_{\mu}^{\textrm{in}}=\boldsymbol{A}_{\mu}-C_{\mu}^{\textrm{in}}\boldsymbol{I}\;.
\end{eqnarray}
Here, $\boldsymbol{A}_{\mu}$ is the electromagnetic vector field and $C_{\mu}^{\textrm{in}}$ is the classical solution to the Maxwell equations with the charged particle source. Therefore, $\boldsymbol{A}_{\mu}^{\textrm{in}}$ satisfies the sourceless Maxwell equations.

After the charged particle passing through the Stern-Gerlach apparatus, the particle is in a spatially separated superposition state, which effectively results in two possible evolutions of the radiation field. The state of the total particle-radiation system is then given by the following form 
\begin{eqnarray}\label{superposition_state_total}
    |\Psi\rangle= \frac{1}{\sqrt{2}}\left(|\psi_L,\uparrow\rangle\otimes |\Psi_L\rangle+|\psi_R,\downarrow\rangle\otimes |\Psi_R\rangle\right)\;,
\end{eqnarray}
where $|\Psi_L\rangle$ and $|\Psi_R\rangle$ are the states of the electromagnetic radiation corresponding to the charged particle states $|\psi_L\rangle$ and $|\psi_R\rangle$. The decoherence due to the electromagnetic radiation is then given by \cite{Danielson:2022tdw}
\begin{eqnarray}
    \mathcal{D}=1-|\langle \Psi_L|\Psi_R\rangle|\;.
\end{eqnarray}

To calculate the overlap between $|\Psi_L\rangle$ and $|\Psi_R\rangle$ of the radiation states, it is convenient to work in the Heisenberg representation. It is assumed that the charge-current $j_i^{\mu} (i=L,R)$ corresponding to the particle state $|\psi_i\rangle$ can be treated as c-number source in Maxwell's equations. In the Heisenberg representation, the radiation field $\boldsymbol{A}_{i,\mu}$ corresponding to $|\psi_i\rangle$ can be given by  
\begin{eqnarray}\label{A_imu_def}
    \boldsymbol{A}_{i,\mu}=\boldsymbol{A}_{\mu}^{\textrm{in}}+G_{\mu}^{\textrm{ret}}(j_i^{\mu})\boldsymbol{I}\;,
\end{eqnarray}
where $G_{\mu}^{\textrm{ret}}(j_i^{\mu})$ is the classical retarded Green function for the Maxwell equations with the source $j_i^{\mu}$. 

After the time at which the particle is recombined, the ``out" radiation field can also be defined analogous to Eq.\eqref{A_in_def} 
\begin{eqnarray}\label{A_out_def}
    \boldsymbol{A}_{i,\mu}^{\textrm{out}}=\boldsymbol{A}_{i,\mu}-C_{\mu}^{\textrm{out}}\boldsymbol{I}\;,
\end{eqnarray}
where $C_{\mu}^{\textrm{out}}$ is the Coulomb field of the charge at late times. From Eqs.\eqref{A_imu_def} and \eqref{A_out_def}, one can get 
\begin{eqnarray}
    \boldsymbol{A}_{i,\mu}^{\textrm{out}}=\boldsymbol{A}_{\mu}^{\textrm{in}}+\left(G_{\mu}^{\textrm{ret}}(j_i^{\mu})-C_{\mu}^{\textrm{out}}\right)\boldsymbol{I}\;.
\end{eqnarray}

By transforming into the Schr\"{o}dinger representation, we have \cite{Gralla:2023oya}
\begin{eqnarray}\label{H_S_trans}
   U_L \boldsymbol{A}_{L,\mu}^{\textrm{out}} U^\dagger_L
   = U_R \boldsymbol{A}_{R,\mu}^{\textrm{out}} U^\dagger_R
   =\boldsymbol{A}_{\mu}^{\textrm{in}}\;,
\end{eqnarray}
where $ U_i$ is the unitary transformation between the Heisenberg and the Schr\"{o}dinger representations. From Eq.\eqref{H_S_trans}, one can get 
\begin{eqnarray}
    D^\dagger  \boldsymbol{A}_{L,\mu}^{\textrm{out}} D= \boldsymbol{A}_{L,\mu}^{\textrm{out}}+\Delta G_{\mu}^{\textrm{ret}} \boldsymbol{I}\;,
\end{eqnarray}
where $D=U_L^\dagger U_R$ can be treated as the displacement operator at late times and $\Delta G_{\mu}^{\textrm{ret}}$ is defined as the difference between $G_{\mu}^{\textrm{ret}}(j_R^{\mu})$ and $G_{\mu}^{\textrm{ret}}(j_L^{\mu})$.

According to the general theory of coherent states \cite{Glauber:1963tx,Zhang:1990fy,2012JPhA...45x4002S}, the radiation states $|\Psi_L\rangle$ and $|\Psi_R\rangle$ are precisely the coherent states. Their overlap can be calculated by using the following expression  
\begin{eqnarray}
    |\langle \Psi_L|\Psi_R\rangle|=|\langle \Psi_0|D|\Psi_0\rangle|
    =\exp\left[ -\frac{1}{2} \langle N\rangle \right]\;,
\end{eqnarray}
with 
\begin{eqnarray}\label{Entangleing_photon}
    \langle N\rangle =\left(K\left[\Delta G_{\mu}^{\textrm{ret}}\right],K\left[\Delta G_{\mu}^{\textrm{ret}}\right] \right),
\end{eqnarray}
where $K\left[\Delta G_{\mu}^{\textrm{ret}}\right]$ represents the positive frequency part associated with the classical solution $\Delta G_{\mu}^{\textrm{ret}}$ and $(,)$ denotes the inner product of the electromagnetic field.

Following DSW, $\langle N\rangle$ is referred as expected number of entangling photons. It is related to the decoherence of the particles state
\begin{eqnarray}
    \mathcal{D}=1-\exp\left[-\frac{1}{2} \langle N\rangle \right]\;.
\end{eqnarray}
If $\langle N\rangle$ is significantly larger than $1$, the superposition state \eqref{superposition_state_total} will be completely decohered. 

In the setup described above, when evaluating the entangling photon number $\langle N\rangle$, the inner product should be computed at a late time, after the particle has been recombined by the experimenter. 
AS done by DSW, we select the future horizon $\mathcal{H}^{+}$ and the future null infinity $\mathcal{F}^{+}$ as the late time Cauchy slice. The inner product between the two electromagnetic fields $A_1$ and $A_2$ is defined in a gauge-invariant and surface-independent manner as \cite{Wald:1995yp,Kay:1988mu}
\begin{align}\label{EM inner product}
    (A_1,A_2) & = -\frac{i}{4\pi}\int_{\Sigma}d^3x \sqrt{h} n^\mu \left(\overline{F}_{1\mu \nu} A_2^\nu-F_{2\mu \nu}\overline{A}_1^{\nu}\right)\;, 
\end{align}
where $F_{\mu\nu}=\nabla_\mu A_\nu-\nabla_\nu A_\mu$ is the electromagnetic field tensor and the overbar denotes the complex conjugate and $n^\mu$ is the future directed unit normal to a spacelike Cauchy surface with the induced metric $h_{\mu\nu}$ and volume element $\sqrt{h} d^3 x$.

Following DSW, we use the horizon-adapted gauge \cite{Danielson:2022tdw,Danielson:2022sga}
\begin{eqnarray}\label{horizon_adapted_gauge}
    A_{\mu} n^{\nu}|_{\mathcal{H}^+}=0.
\end{eqnarray}
Then the inner product evaluated on the future horizon $\mathcal{H}^+$ can be written as \cite{Gralla:2023oya,Kay:1988mu}
\begin{align}\label{EM horizon inner product}
    (A_1, A_2&)_{\mathcal{H}^+} = \frac{1}{4\pi^2} \int_{S^2} r_+^2 d\Omega \int_{0}^{+\infty} \omega d\omega   q^{AB} \overline{A_{1,A}(\omega,x^A)} A_{2,B}(\omega,x^A)\;,
\end{align}
where $q^{AB}$ is the inverse metric of the transverse two sphere $S^2$ on the horizon and $A_{A}(\omega,x^A)$ is the Fourier transform of the electromagnetic field $A_{A}(V,x^A)$ on the horizon.

By a simple estimation, it can be shown that the decoherence from the future null infinity is zero. This is caused by the fact that the decay of the gauge potential asymptotically like $\frac{1}{r}$ in the spherically symmetric spacetime \cite{Danielson:2021egj}. Then the contribution from the future null infinity to the entangling photon number is asymptotical $\frac{1}{r^2}$, which is zero when taking the limit $r\rightarrow +\infty$. Therefore, to evaluate the entangling photon number, one can safely calculate the inner product on the future horizon. In the following, our aim is to evaluate the expected number of entangling photons at late times for the static charged particle in the {\rn} black hole.

\section{Estimation of entangling photon number in {\rn} black holes} 
\label{sec:est_num}

The {\rn} black hole is described by the metric 
\begin{eqnarray}
 ds^2=-f(r)dt^2+\frac{1}{f(r)}dr^2+r^2d\Omega^2\;,
\end{eqnarray}
where the blacken factor $f(r)$ is given by 
\begin{eqnarray}
    f(r)=1-\frac{2M}{r}+\frac{Q^2}{r^2}\;,
\end{eqnarray}
with $M$ and $Q$ being the mass and the charge of black hole. 

When $M>Q$, the two roots of the equation $f(r)=0$ gives the inner and the outer horizons, which are given by 
\begin{eqnarray}
    r_{\pm} = M\pm \sqrt{M^2-Q^2}\;.
\end{eqnarray}
In the extremal case $M=Q$, the two horizons coincide.

It is clear the metric in Schwarzschild-like coordinates are not regular on the horizon. To eliminate the coordinate singularity at $r=r_+$, one can introduce the Kruskal-like coordinates as 
\begin{eqnarray}\label{Kruskal_coord}
    U= - e^{-\kappa_{+}(t-r^*)}\;,\;\;\;
    V= e^{\kappa_{+}(t+r^*)}\;,
\end{eqnarray}
where the tortoise coordinate $r^*$ is defined as 
\begin{eqnarray}
    r^*=r+\frac{1}{2\kappa_+}\ln \left| \frac{r-r_+}{r_+} \right|
    +\frac{1}{2\kappa_-}\ln \left| \frac{r-r_-}{r_-} \right|\;,
\end{eqnarray} 
and the surface gravity $\kappa_{\pm}$ on the horizons are given by
\begin{eqnarray}
    \kappa_{\pm} = \frac{r_{\pm} - r_{\mp}}{2r_{\pm}^2}\;.
\end{eqnarray}

Using the Kruskal-like coordinates \eqref{Kruskal_coord}, the {\rn} metric has the form of 
\begin{eqnarray}
    ds^2=-\frac{r_+r_-}{\kappa_+^2 r^2} \left( \frac{r-r_-}{r_-}\right)^{1+r_-^2/r_+^2}e^{-2\kappa_+ r}dUdV +r^2 d\Omega^2\;,
\end{eqnarray}
where $r$ should be understood as the function of the Kruskal coordinates $U$ and $V$ towards the following relation 
\begin{eqnarray}
    UV=-\left( \frac{r-r_+}{r_+}\right)\left( \frac{r_-}{r-r_-}\right)^{r_-^2/r_+^2} e^{2\kappa_+ r}\;.
\end{eqnarray}
Note that on the future Killing horizon $\mathcal{H}^{+}$, $U=0$, $V>0$, $V$ is the affine parameter, and $n^\mu=\left(\frac{\partial}{\partial V}\right)^\mu$ is the null generator of the Killing horizon \cite{Townsend:1997ku}.

We now calculate the entangling photon number \eqref{Entangleing_photon} in terms of the Eq.\eqref{EM horizon inner product}, which reads
\begin{eqnarray}\label{entangle_G_exp}
    \langle N\rangle = \frac{1}{4\pi^2}  \int_{S^2} r_+^2 d\Omega \int_{0}^{+\infty} \omega d\omega q^{AB} \overline{\Delta G^{\textrm{ret}}_{A}(\omega,x^A)}  \Delta G^{\textrm{ret}}_{B}(\omega,x^A)\;.
\end{eqnarray}
The retarded Green function $G_{\mu}^{\textrm{ret}}(j_i^{\mu})$ for the Maxwell equations with the source $j_i^{\mu}$ can be approximated by the static solution with the static source $j_i^{\mu}$ discussed above.

Note that the static source is maintained only for a affine time $\Delta V$. When $\Delta V$ is sufficiently large, the Fourier transform of the difference between the two Green functions can be well approximated as 
\begin{eqnarray}\label{Green_Fourier}
    |\Delta G^{\textrm{ret}}_{\theta}(\omega,x^A)|=\frac{1}{\omega} \Delta A_{\theta}  \;. 
\end{eqnarray}
In the next section, we will demonstrate that, in our setup, only the $\theta$ component of the Green function is nonvanishing in the horizon-adapted gauge. It is clear that there is a logarithmic divergence at $\omega=0$. This can be fixed by introducing a low frequency cutoff $\omega\sim 1/\Delta V$ \cite{Danielson:2024yru}. By using Eq.\eqref{Green_Fourier} to evaluate the entangling photon number in Eq.\eqref{entangle_G_exp}, one can get 
\begin{eqnarray}
    \langle N\rangle \approx C \log \Delta V\;,
\end{eqnarray}
where the decohering flux $C$ is given by 
\begin{eqnarray}\label{decohering_flux}
   C=\frac{1}{4\pi^2} \int \left|\Delta A_{\theta}\right|^2 d\Omega\;. 
\end{eqnarray}
Note that for large $\Delta V$, the contribution to the entangling photon number is mainly come from modes with the frequency $1/\Delta V$. 
For this reason, it argued that the quantum superposition is decohered by the soft photon radiated from the charged particle.

According to the relation between the affine time $V$ and the Killing time $t$, which is given by $V=e^{\kappa_+ t}$, one can further get the entangling photon number in the form of 
\begin{eqnarray}\label{linear_rel}
    \langle N\rangle \approx C \kappa_+ \Delta T\;,
\end{eqnarray}
which means the entangling photon number is linearly growing with the Killing time. This results indicates that the quantum superposition state hold by the experimenter will be decohered after sufficiently long time by radiation the low frequency soft photon into the black hole horizon. However, in the extremal black hole limit, where $\kappa_+\rightarrow 0$, the entangling photon number will be sufficiently suppressed and the coherence of the spatial separated quantum superposition will be maintained. Thus far, our final task is to compute the exact expression of the decohering flux given in Eq.\eqref{decohering_flux} for the {\rn} black holes.

\section{Calculation of decohering flux}
\label{sec:decoh_flux}

To compute the exact expression of the decohering flux,
we assume that the left and the right branches of the charge currents persist forever. Then the left and the right branches of the charge current corresponding to the spatial quantum superposition state can be treated as the static sources. Therefore, we firstly study the static electromagnetic field generated by the classical static charged source in {\rn} black hole.

The charge-current of a point source in a curved spacetime is given by
\begin{align}\label{EM source}
    J^\mu = \int q u^\mu \frac{\delta^{(4)}(x- x(\tau))}{\sqrt{-g}} d\tau
\end{align}
where $q$ is the electric charge and $u^{\mu}=\frac{dx^{\mu}}{d\tau}$ is the four velocity of the charged particle. For a static charge on $z$ axis of a {\rn} black hole, we have
\begin{align}\label{Eq_source}
    J = \frac{q}{r^2} \delta(r-r_0)\delta(1-\cos\theta) \frac{\partial}{\partial t}
\end{align}
where $r_0$ is the location of the charged particle.

The regular solution with charge $q$ at rest has been found in \cite{Leaute:1976sn}. It is shown that the nonvanishing component of the electromagnetic potential is $A_t$. The equation for $A_t$ is given by 
\begin{eqnarray}
    \frac{\partial}{\partial r} \left(r^2 \frac{\partial A_t}{\partial r} \right) +\frac{1}{f(r)}\left[ \frac{1}{\sin\theta}  \frac{\partial}{\partial \theta} \left(\sin\theta\frac{\partial A_t}{\partial \theta}\right)+\frac{1}{\sin^2\theta} \frac{\partial^2 A_t}{\partial\phi^2} \right]=-r^2 J^{t}
\end{eqnarray}
In fact, for the source given in Eq.\eqref{Eq_source}, the electromagnetic potential $A_t$ is axisymmetric, i.e. $A_t=A_t(r,\theta)$. The solution to the equation can be given in a closed form as 
\begin{eqnarray}\label{At_sol}
    A_t=\frac{q}{r_0 r}\left[M+\frac{(r-M)(r_0-M)-(M^2-Q^2)\cos\theta}{\left[ (r-M)^2+(r_0-M)^2-2(r-M)(r_0-M)\cos\theta - (M^2-Q^2)\sin^2\theta\right]^{1/2}} \right]
\end{eqnarray}
It is easy to see that the solution is also valid for the extremal {\rn} black hole with $M=Q$. In this case the Maxwell equation \eqref{Eq_source} is formally identical to Laplace’s equation in Minkowski space.

It should be noted that the gauge potential given in Eq.\eqref{At_sol} is not consistent with the horizon-adapted gauge given in Eq.\eqref{horizon_adapted_gauge}. We apply the method in \cite{Gralla:2023oya} to compute the expression of the electromagnetic field in the horizon-adapted gauge. One can firstly compute the gauge-invariant field strength and then reconstruct a suitable $A_A^c$ by solving the Maxwell equations on the horizon.

It is easy to show that in the horizon-adapted gauge, the Maxwell equation on the horizon can be written as \cite{Gralla:2023oya} 
\begin{eqnarray}\label{Max_eq_horizon}
    q^{AB} \nabla_A A_B&=&\frac{1}{2} \epsilon^{AB} ~^*F_{AB}\;, \label{Max_eq_horizon_1}\\
    \epsilon^{AB} \nabla_A A_B&=&\frac{1}{2} \epsilon^{AB} F_{AB}\;, \label{Max_eq_horizon_2}
\end{eqnarray}
where $x^{A}=\{\theta,\phi\}$ is the transverse coordinates, $q_{AB}=\textrm{diag}\{r_+^2,r_+^2\sin^2\theta\}$ is the transverse metric on the horizon, $\epsilon_{AB}$ is the two-dimensional area element, and $~^*F_{AB}$ is the dual electromagnetic tensor. It should be noted that the right hand sides of Eq.\eqref{Max_eq_horizon_1} and Eq.\eqref{Max_eq_horizon_2} are just the radial electric and magnetic fields, which are clearly gauge invariant.

For the {\rn} black hole spacetime, the Maxwell equations on the horizon $r=r_+$ are explicitly given by 
\begin{align}
    \frac{1}{\sqrt{q}}\partial_\theta\left(\sqrt{q}q^{\theta\theta} A^c_\theta \right) & = \frac{1}{2} \epsilon^{AB} \ \! {}^*\!F^c_{AB} \;, \label{horizon_eq1} \\
        \frac{1}{\sqrt{q}}\partial_\theta A^c_{\phi}& = \frac{1}{2}\epsilon^{AB} F^c_{AB} \label{horizon_eq2}\;.
\end{align}
By assuming that $A_A^c$ is independent of the angular coordinate $\phi$, the integrals can be done in closed form. After some calculation (see Appendix \ref{appendix}), one can find the nonvanishing component of the ``Coulomb" gauge potential on the horizon as
\begin{align}\label{A_horizon}
    A_\theta^c= \frac{q r_+ (r_+-M)\sin\theta}{r_0 \left[r_0-M-(r_+-M) \cos \theta\right]}\;,
\end{align}
where an integral constant $\frac{qr_+}{r_0}\frac{r_0-M}{r_+-M}$ has been selected to eliminate the divergence on the pole. The above equation \eqref{A_horizon} refers to the Coulomb field generated by a single static point charge $q$ in the {\rn} black hole spacetime. It is clear that the gauge potential vanishes at the extremal limit $r_+\rightarrow M$, which is an indication of the black hole Meissner effect \cite{Wald:1974np,King:1975tt,Bicak:1980du,Bini:2008iwy}.

For the superposition experiment, the experimenter prepares the charged particle at in a superposition of the position eigenstates. The two eigenvalues are two nearby radii $r_0 \pm \epsilon/2$, where $\epsilon$ denotes the distance between the two spatial positions. To calculate the decohering flux $C$ given by Eq.\eqref{decohering_flux}, we need the difference of the two vector potentials corresponding the two different sources. Since the proper distance between two sources is $\sqrt{g_{rr}}|_{r=r_0} \epsilon$, the combination of the left and the right sources is effectively equivalent to a electric dipole with the moment $p = q \sqrt{g_{rr}}|_{r=r_0} \epsilon$. By keeping the moment $p$ fixed, in the limit $\epsilon \to 0$, the difference between the left and the right gauge potential is given by 
\begin{align}
    \Delta A_\theta= A^{\rm stat}_{R,\theta} - A^{\rm stat}_{L,\theta} = \frac{p}{q \sqrt{g_{rr}}|_{r=r_0}} \left.\frac{\partial A_\theta^c}{\partial r_0}\right|_{q=1}.
\end{align}
It is found that 
\begin{eqnarray}
    \Delta A_\theta=  -\frac{p r_+ ((r_+-M)) \left(M-2r_0+(r_+-M)\cos \theta \right) \sin\theta}{r_0^2 \sqrt{g_{rr}}|_{r=r_0}\left[r_0-M-(r_+-M) \cos \theta\right]^2} \;.
\end{eqnarray}

Finally, one can evaluate the decohering flux integral in Eq.\eqref{decohering_flux}. It is found that 
\begin{eqnarray}\label{C_RN}
     C &=& \frac{p^2 r_+^2 (r_0^2-2Mr_0+Q^2)}{2\pi r_0^6}\left[\frac{2 M}{ r_+-M} \ln \frac{r_0-r_+}{r_0+r_+-2 M}\right.\nonumber\\
   &&\left.+\frac{4 \left[M r_0^2 \left(3 r_0-5M\right)+r_+ r_0 \left(r_+-2 M\right) \left(4 r_0-9M\right)-3 r_+^2 \left(r_+-2 M\right)^2\right]}{3  \left(r_0-r_+\right)^2  \left(r_0+r_+-2 M\right)^2}\right]\;.
\end{eqnarray}
This result along with Eq.\eqref{linear_rel} gives the exact expression for the expected number of the entangling photon generated by the charged particle in the spatial separated quantum superposition state in the {\rn} black hole.

\begin{figure}
  \centering
  \includegraphics[width=8cm]{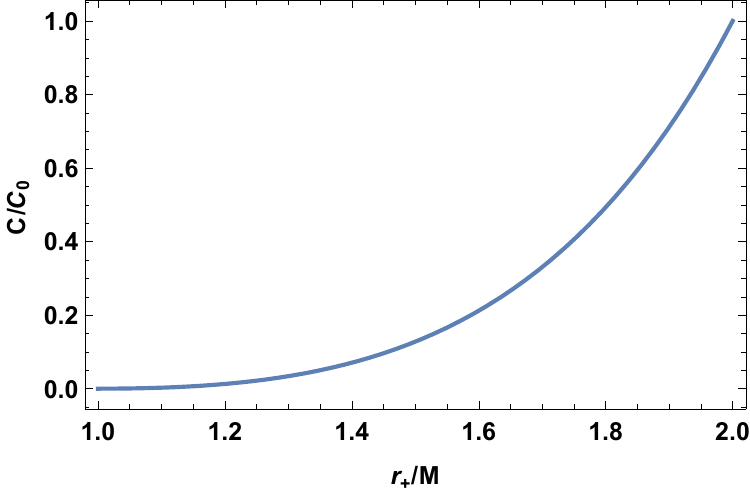}
  \caption{Decohering flux $C$ plotted as the function of $r_+/M$, where $r_0=3M$ and $C_0$ is the value of decohering flux at $Q=0$. In the extremal limit $r_+=M$, the decohering flux vanishes due to the black hole Meissner effect. }
  \label{decohering_flux_plot}
\end{figure}

It is easy to check that the decohering flux vanishes in the extremal limit $r_+=M$. The decohering flux $C$ as the function of $r_+/M$ is plotted in Figure \ref{decohering_flux_plot}, which shows the exact behavior in the extremal limit. This phenomenon is related to the black hole Meissner effect, which refers to the fact that extremal black holes tend to expel magnetic and electric fields. In the extremal black hole case, the radiated photon cannot propagate into the black hole horizon. The ``which way" information of the superposed particle which carried by the soft photons cannot be captured by the black hole. Therefore, the decoherence effect of the extremal black hole is completely suppressed and the coherence of the quantum superposition is maintained.

\section{Comparing the decoherence of {\rn} black holes with ordinary material systems}
\label{Sec_comparing}

Biggs and Maldacena have found that the decoherence effect of a black hole is comparable to that of an ordinary material system \cite{Biggs:2024dgp}. In this section, let us address the question of whether the decoherence caused by a {\rn} black hole is equivalent to that of an ordinary material system with the same temperature and charge.

To make a qualitative comparison, we consider the case where the charged particle is placed far from the horizon. In this scenario, the decohering flux presented in Eq.\eqref{C_RN} can be approximated in the large $r_0$ limit as
\begin{eqnarray}
   C = \frac{8p^2(M^2-Q^2)r_+^2}{3\pi r_0^6}\;.
\end{eqnarray}
Then the entangling photon number is approximated as 
\begin{eqnarray}
    \langle N \rangle = \frac{8p^2(M^2-Q^2)^{3/2}}{3\pi r_0^6}\Delta T\;,
\end{eqnarray}
and the constant decohering rate can be given by 
\begin{eqnarray}\label{Decohering_rate}
    \Gamma =  \frac{8p^2(M^2-Q^2)^{3/2}}{3\pi r_0^6}\;.
\end{eqnarray}

We first make a qualitative comparison with the local calculation of decohering effect presented in \cite{Biggs:2024dgp}. It was shown that the decohering rate can be given by 
\begin{eqnarray}
    \Gamma\sim \frac{p^2}{r_0^6} S(\omega=0)\;,
\end{eqnarray}
where $S(\omega)$ is the Wightman function. For the electromagnetic case, the Wightman function is related to the classical absorption cross section by the following relation \cite{Biggs:2024dgp}
\begin{eqnarray}
    S(\omega)\sim T_H \sigma_{\textrm{abs}}\;,
\end{eqnarray}
where $T_H=\kappa_+/2\pi$ is the Hawking temperature of the outer horizon. The absorption cross section of the electromagnetic waves for the {\rn} black hole was computed by Gubser in \cite{Gubser:1997cm} (see also \cite{Crispino:2009zza}), which is given by 
\begin{eqnarray}
    \sigma_{\textrm{abs}}= \frac{(r_+-r_-)^3}{3T_H}\left(1+\frac{\omega^2}{4\pi^2 T_H^2}\right)\;.
\end{eqnarray}
It is straightforward to verify that the result from the local calculation is consistent with our result in Eq.\eqref{Decohering_rate}.

According to the fluctuation
dissipation theorem \cite{Kubo:1966fyg}, the Wightman function is related to the response function $\chi(\omega)$ in the low frequency limit by the following relation 
\begin{eqnarray}
   \textrm{Im} \chi(\omega)\sim \frac{1}{2}\beta \omega S(\omega)\;,
\end{eqnarray}
where $\beta$ is the inverse Hawking temperature. It is well known that the response function $\chi(\omega)$ describes the expectation value of an operator due to a small perturbation by an external source that couples to that operator. For the {\rn} black hole interacting with an external electric field, one can get the response function as 
\begin{eqnarray}\label{resp_bh}
     \textrm{Im} \chi_{\textrm{BH}}(\omega)\sim \omega  r_+^2 (M^2-Q^2)\sim \omega r_+^4 \left(1-\frac{2Q^2}{r_+^2}\right)\;,
\end{eqnarray}
where in the last step we have used the approximation condition $Q/r_+\ll 1$.

We now compare this with the response function of a solid conducting sphere with radius $R$, charge $Q$ and resistivity $\rho$. When the conducting sphere is placed in a static external electric field, the free charge $Q$ and the induce charge $Q_i$ will rearrange to cancel the electric field inside the sphere. The dipole moment due to the induced charge $Q_i$ can be approximated as $P\sim Q_i R$. However, we are mainly concerned with the case where the external electric field $\vec{E}_{\textrm{ext}}$ oscillates with the frequency $\omega$. In this scenario, the cancellation is incomplete due to the finite resistivity, which prevents charges from rearranging instantaneously \cite{Biggs:2024dgp}. We can get 
\begin{eqnarray}\label{current_eq}
    \vec{E}_{\textrm{ext}}-\vec{E}_i=\rho\vec{J}\;,
\end{eqnarray}
where $E_i$ is the electric field generated by the induced charge and $\vec{J}$ is the current density. Here, we ignore the electric field generated by the free charge $Q$.

The current density $\vec{J}$ can be decomposed into two parts, $\vec{J}_Q$ and $\vec{J}_i$, which correspond to the currents due to the movement of the free charge $Q$ and the induced charge $Q_i$ under the external electric field. The current density $\vec{J}_i$ can be related to the dipole moment via the relation $\int d^3x \vec{J}=\frac{d\vec{P}}{dt}$. This leads to the expression $J_i\sim \frac{-i\omega P}{R^3}$.

To estimate the current density $J_Q$, we recall the definition of the current density as 
\begin{eqnarray}
    J_Q=\frac{\Delta Q}{\Delta S \Delta t} \sim \frac{i\lambda\omega Q^2 E_{\textrm{ext}}}{R^2} \;,
\end{eqnarray}
where $\lambda$ is a numerical factor that may be determined by comparing with the black hole case. In this estimation, we have used the fact that the free charge $\Delta Q$ passing through the area element $\Delta S$ in the time interval $\Delta t$ should be proportional to the total free charge $Q$ and the electric field force $Q E_{\textrm{ext}}$. Substituting these into Eq.\eqref{current_eq}, one can obtain 
\begin{eqnarray}
    E_{\textrm{ext}}\left(1-\frac{i\lambda\omega\rho Q^2}{R^2}\right)\sim \frac{P}{R^3} \left(1-i\omega \rho \right)\;.
\end{eqnarray}
Then, we have 
\begin{eqnarray}\label{resp_cond}
    \textrm{Im} \chi_{\textrm{cond}}(\omega)\sim \omega \rho  R^3 \left(1-\frac{\lambda Q^2}{R^2}\right)\;.
\end{eqnarray}

We assume that the solid conducting sphere has the same size as the black hole, i.e. $R\sim r_+$. Compare Eq.\eqref{resp_cond} with the response function of the {\rn} black hole given in Eq.\eqref{resp_bh}, we can see that $\rho/R$ should be of order unity. If $Q/R\ll 1$, the numerical factor $\lambda$ can be fixed to be $2$. Therefore, we can conclude that the decoherence effect caused by the {\rn} black hole is equivalent to that of an ordinary matter system with the same size and charge. This observation extends the conclusion of \cite{Biggs:2024dgp}, which suggests that the decoherence due to a black hole horizon is equivalent to that of any other quantum system with the same temperature, to the charged case.

\section{Conclusion and discussion}
\label{sec:con_disc}

In summary, we have studied the DSW decoherence effect for a static charged particle in the {\rn} black holes. The explicit expression for the decoherence is obtained by evaluating the expected number of the entangling photon at late times. Our calculation is limited to the case of electromagnetic radiation from the charged particle in a spatial separated quantum superposition state. It is shown that the quantum superposition is decohered by the low frequency photons that propagate through the black hole horizon. We also discuss the case of the extremal {\rn} black hole. It is shown that the decoherence of quantum superposition is completely suppressed due to the black hole Meissner effect. We also compare the decoherence effect due to the {\rn} black hole with that of the ordinary material system and find that the decoherence effect caused by the {\rn} black hole is equivalent to that of an ordinary matter system with the same size and charge.

It is expected the calculation can be generalized to the gravitational case. However, it is a more challenging task because the gravitational radiation generated by the static gravitational source in the background of the curved spacetime should be taken into account. This will be a subject of future investigation.

\appendix
\section{Calculation of ``Coulomb" gauge potential on the horizon}
\label{appendix}

The ``Coulomb" gauge potential on the horizon, given by Eq.\eqref{A_horizon}, is obtained by solving the Maxwell equations on the horizon, Eq.\eqref{horizon_eq1} and Eq.\eqref{horizon_eq2}. Since the static charge is placed on the $z$ axis of a {\rn} black hole, the``Coulomb" gauge potential on the horizon is independent of the angular coordinate $\phi$. It is evident that the right-hand sides of Eq.\eqref{horizon_eq1} and Eq.\eqref{horizon_eq2} are gauge invariant. Thus, they can be calculated by using the gauge potential expressed in Eq.\eqref{At_sol}. However, this gauge potential have only the $t$ component. So the right-hand side of Eq.\eqref{horizon_eq2} is zero, which gives the fact that $A^c_{\phi}$ is a constant. Without loss of generality, we can set this constant to zero. The only non-vanishing component of the ``Coulomb" gauge potential $A^c_A$ on the horizon is $A_\theta^c$.

For the right-hand side of Eq.\eqref{horizon_eq1}, it can be shown that 
\begin{eqnarray}
    \frac{1}{2} \epsilon^{AB} \ \! {}^*\!F^c_{AB}=\frac{1}{\sqrt{q}}{}^*\!F_{\theta\phi}=\frac{\sqrt{-g}}{\sqrt{q}}F^{tr}=\partial_r A_t\;.
\end{eqnarray}
By evaluating the derivative of $A_t$ on the horizon, one can get
\begin{eqnarray}
    \frac{1}{2} \epsilon^{AB} \ \! {}^*\!F^c_{AB}=-\frac{q  (r_+-M)\left[(r_+-M)\left(\cos^2\theta +1\right) -2 (r_0-M) \cos\theta \right]}{r_0 r_+ \left[r_0-M-(r_+-M) \cos \theta\right]^2}\;.
\end{eqnarray}
Finally integrating Eq.\eqref{horizon_eq1}, one can get 
\begin{eqnarray}
    A_\theta^c= \frac{q r_+ (r_+-M)\sin\theta}{r_0 \left[r_0-M-(r_+-M) \cos \theta\right]}-\frac{q r_+ (r_0-M)}{r_0 (r_+-M) \sin\theta}+\frac{\lambda}{\sin\theta}\;,
\end{eqnarray}
where $\lambda$ is a integral constant. It is obvious that the divergence of $A_\theta^c$ on the pole $\theta=0,\pi$ can be eliminated by selecting the integral constant $\lambda$ to be $\frac{q r_+ (r_0-M)}{r_0 (r_+-M)}$.

\acknowledgments  The authors would like to thank Hongji Wei for helpful
discussions.


\bibliographystyle{JHEP}
\bibliography{biblio.bib}
\end{document}